\documentclass[12pt]{iopart}

\usepackage{graphicx}

\begin{document}

\title{Characterization of the (Cu,C)Ba$_2$Ca$_3$Cu$_4$O$_{11+\delta}$ single crystals grown under high pressure}

\author{Chengping He, Xue Ming, Jin Si, Xiyu Zhu, Jinhua Wang, and Hai-Hu Wen$^{*}$}

\address{Center for Superconducting Physics and Materials, National Laboratory of Solid State Microstructures and Department of Physics, Collaborative Innovation Center of Advanced Microstructures, Nanjing University, Nanjing 210093, China}
\ead{hhwen@nju.edu.cn}
\vspace{10pt}
\begin{indented}
\item[]November 2021
\end{indented}

\begin{abstract}
By using high pressure and high temperature (3.7 GPa, 1120 $^{\circ}$C) synthesis technique, we have grown (Cu,C)Ba$_2$Ca$_3$Cu$_4$O$_{11+\delta}$ single crystals. X-ray diffraction, scanning electron microscopy, resistivity and magnetization measurements are carried out and all show that the samples have good quality. The single crystal has onset and zero-resistance transition temperatures of about 111 K and 109.6 K, indicating a very narrow transition width, which is consistent with a rather sharp magnetization transition. Magnetization hysteresis loops (MHLs) are also measured, showing a pronounced second peak effect in the intermediate temperature region. The magnetic critical current density calculated from the MHLs at 77 K and 1.5 T is about 6.4$\times$10$^4$ A/cm$^2$. By using a criterion of 1$\%$ normal state resistivity, we have determined the irreversibility line which exhibits an irreversibility field of about 8 T at 77 K. Compared with other layered systems, it is easy to find that the irreversibility line is rather high and could be further improved with the optimized transition temperature of about 118 K as previously discovered in polycrystalline samples.
\end{abstract}

\noindent{\it Keywords}: superconductor, (Cu,C)-1234 single crystal, application
%\maketitle

\section{Introduction}

The discovery of high temperature superconductivity in La-Ba-Cu-O in 1986 \cite{LaBaCuO} has dramatically promoted scientific research and applications of superconductivity. Since then, extensive efforts have been made to explore other new cuprate superconductors with superior properties for practical applications. For that purpose, high critical transition temperature (\emph{T}$_c$), high critical current density (\emph{J}$_c$) and high irreversibility field (\emph{H}$_{irr}$) are three key factors. But of course, nontoxic elements are more welcome in the industry applications. The irreversibility line \emph{H}$_{irr}$(\emph{T}) is defined as the boundary between the magnetically irreversible region with finite \emph{J}$_c$ and the reversible region with zero \emph{J}$_c$. The practical physical meaning of \emph{H}$_{irr}$(\emph{T}) is that it determines the upper limit of the capability for a superconductor to carry nondissipative supercurrent \cite{Hirr}. Among cuprate superconductors, the application of Bi-based compounds is limited by the low \emph{H}$_{irr}$ at liquid nitrogen temperature due to the high anisotropy \cite{Bi-based}. For the Tl- and Hg-based compounds, the toxic Tl and Hg elements as well as the high anisotropy constrain their applications \cite{Tl-based1, Tl-based2, Hg-based}. Furthermore, the nontoxic Y-based superconductor YBa$_2$Cu$_3$O$_{7-\delta}$ (YBCO) has a high \emph{T}$_c$ ($\sim$90 K) and a high \emph{H}$_{irr}$ ($\sim$10 T at 77 K) \cite{YBCO1, YBCO2}, which indicates great potential for applications. However, the obstacles to the widespread application of YBCO are to produce it in wire form in long length and at price competitive to copper \cite{YBCO application}.

In order to avoid the toxic elements like Tl and Hg, some efforts in synthesis of nontoxic superconducting compounds have been made. One of them is the phase (Cu,C)Ba$_2$Ca$_3$Cu$_4$O$_{11+\delta}$, namely (Cu,C)-1234, which was first reported in 1994 \cite{CuC-kawashima1994}. The (Cu,C)-1234 superconductor has a high \emph{T}$_c$ of 118 K even in the highly overdoped state \cite{CuC 118K}. Some discussions were made on the anistropy of the system based on the estimates of the penetration depth and coherence length along c-axis, but this lacks an experimental verification \cite{CuMg, CuC anisotropy}. At zero applied magnetic field, the value of \emph{J}$_c$ reaches 6$\times$10$^6$ A/cm$^2$ at 4.2 K and 6.5$\times$10$^5$ A/cm$^2$ at 77 K, respectively. And most strikingly, it holds the record of the highest irreversibility line among all superconductors in the liquid nitrogen temperature region. The irreversibility field reaches 15 T at 86 K and 5 T at 98 K, respectively \cite{CuC final}. Furthermore, the (Cu,C)-1234 thin films have also been fabricated successfully \cite{CuC film1}. Based on these excellent characteristics, the (Cu,C)-1234 superconductor is considered to be a promising candidate for practical applications above liquid nitrogen temperature.

However, these works were carried out on polycrystalline samples or thin films. In order to avoid the influence of defects and grain boundaries in polycrystalline samples and to reveal the intrinsic properties of this material, it is highly desirable to prepare high-quality single crystals. For example, the critical current density in polycrystals is limited by the weak-links between grains \cite{defects}, while this problem can be solved in single crystals. Crabtree et al. \cite{YBCO current two times} have found that the critical current density in single crystals YBCO is two orders of magnitude greater than that in polycrystalline samples. In addition, Dubois et al. \cite{IL high} have found that there is an enhancement of irreversibility line in YBCO single crystal compared with that in YBCO ceramics. Note that Tokiwa et al. \cite{Cu-1234 single crystal} have grown the CuBa$_2$Ca$_3$Cu$_4$O$_y$ single crystals before. However, there was no consensus on whether the samples contained carbon at that time and the maximum \emph{T}$_c$ is 105 K, which is significantly lower than the value of 118 K in the polycrystalline samples.

In the present work, we report the growth and characterization of high-quality (Cu,C)-1234 single crystals. Magnetic and resistive measurements both show sharp superconducting transitions with \emph{T}$_c$ of about 111 K. A pronounced second peak effect is observed in the intermediate temperature region from MHLs and \emph{J}$_c$(\emph{H}) curves. The \emph{J}$_c$ calculated from the MHLs at 77 K and 1.5 T with the Bean critical state model is about 6.4$\times$10$^4$ A/cm$^2$. By using a criterion of 1$\% \rho_n$ with $\rho_n$ the normal state resistivity, we find a rather high irreversibility line of the single crystal compared with other systems, especially above 77 K. However, the irreversibility line of the single crystals is slightly lower than that previously reported in the (Cu,C)-1234 polycrystalline samples with \emph{T}$_c$ of 116 K \cite{CuC final}, which is probably due to the lower \emph{T}$_c$ of the present single crystals.

\section{Experimental details}

The (Cu,C)-1234 single crystals were grown under high pressure and high temperature by self-flux method. First, the precursor BaCuO$_{2+x}$ was prepared by calcining the well ground mixture of BaO$_2$ (99$\%$, Aladdin) and CuO (99.995$\%$, Alfa Aesar) in flowing oxygen at 900 $^{\circ}$C for 60 h with several intermediate grindings. Another precursor Ca$_2$CuO$_3$ was prepared first by calcining the well ground mixture of CaCO$_3$ (99.99$\%$, Alfa Aesar) and CuO at 950 $^{\circ}$C for 20 h in air, and then reheated in flowing oxygen at 900 $^{\circ}$C for 40 h with several intermediate grindings. Subsequently, the BaCuO$_{2+x}$, Ca$_2$CuO$_3$, CuO, CaCO$_3$, BaCO$_3$ (99.997$\%$, Alfa Aesar) and appropriate amount of Ag$_2$O (used as oxidizer) (99+$\%$, Alfa Aesar) were weighed and ground thoroughly in an agate mortar with the nominal composition of (Cu$_{0.6}$C$_{0.4}$)Ba$_2$Ca$_3$Cu$_4$O$_{11+\delta}$. The mixture was pressed into pellets and sealed in a gold capsule. These preparation procedures were carried out in a glove box filled with purified argon. For the high-pressure and high-temperature synthesis, the gold capsule was placed in a BN capsule which was surrounded by a graphite tube heater, heated up to 1120 $^{\circ}$C and kept at this temperature for 4 h under the pressure of 3.7 GPa in a piston-cylinder high pressure apparatus (LP 1000-540/50, Max Voggenreiter). After that, the sample was cooled to room temperature in 5 minutes before the pressure was released.

The crystal structure of the single crystals was examined by powder X-ray diffraction (XRD, Bruker D8 Advance) with Cu-K$_{\alpha}$ radiation and the PowderX software was used to analyze the raw XRD data. The chemical composition of the single crystals was analyzed by the energy dispersive X-ray spectrum (EDS) attached to a scanning electron microscopy (SEM, Phenom Prox) and electron probe microanalyzer (EPMA-8050G, Shimadzu). The dc magnetization measurements were carried out on a vibrating sample magnetometer (PPMS-VSM 9 T, Quantum Design) with magnetic fields up to 7 T. The resistivity was measured by the standard four-probe method using the physical property measurement system (PPMS-16 T, Quantum Design) with magnetic field ranging from 0 to 11 T. For both the magnetic and resistive measurements, the magnetic field was always parallel to c-axis of the single crystal.

\section{Results and discussion}

\begin{figure}
  \centering
  \includegraphics[width=4in]{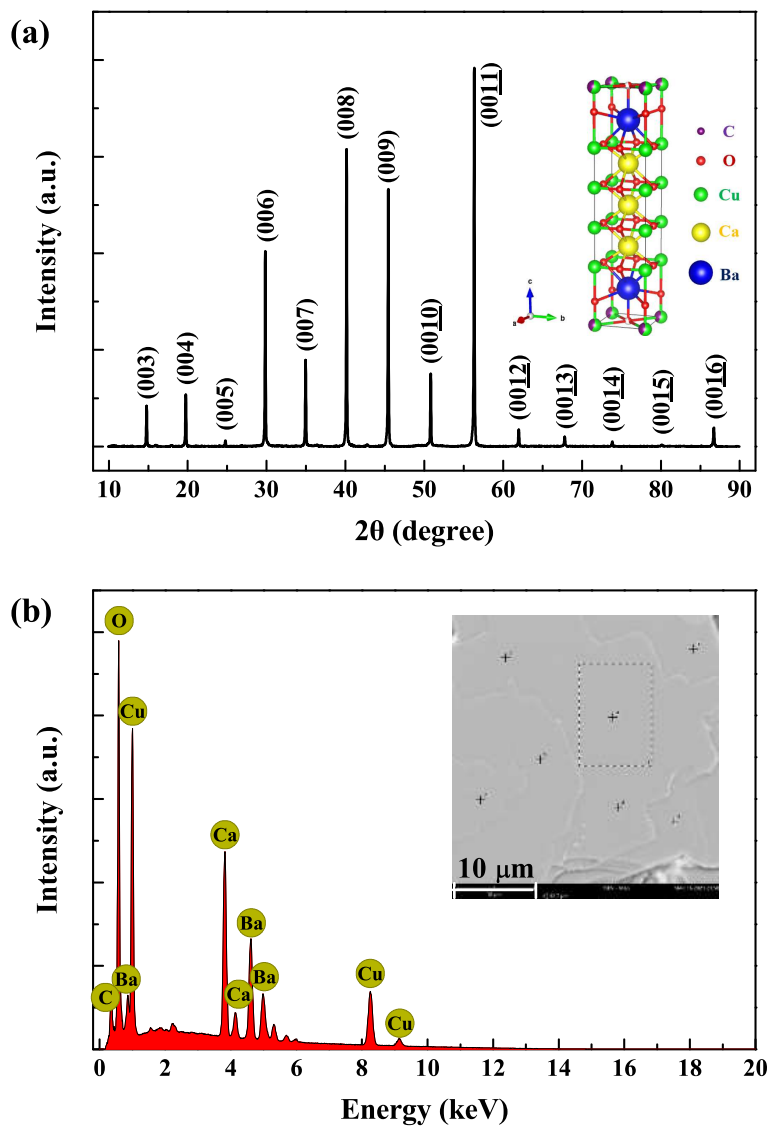}
\caption{\label{fig1}(a) XRD pattern for the (Cu,C)-1234 single crystal. The inset shows the schematic crystal structure of (Cu,C)-1234. (b) A typical EDS spectrum for the single crystal. The inset shows the SEM image of the single crystal. The dashed frame and the crosses show the places where the compositions are analyzed.}
\end{figure}

Figure~\ref{fig1}(a) shows the XRD pattern for the (Cu,C)-1234 single crystal. All diffraction peaks can be well indexed with a tetragonal structure (space group: \emph{P4/mmm}). Only sharp peaks along (00l) orientation can be observed. The full width at half maximum (FWHM) of each peak is about 0.10$^{\circ}$, which indicates high c-axis orientation and good crystalline quality of the single crystal. The lattice constant c analyzed by PowderX software is 17.96 $\mathring{A}$, which is consistent with the previous results on polycrystalline samples \cite{CuC-kawashima1994, crystal structure}. The inset in Fig.~\ref{fig1}(a) shows the schematic crystal structure of (Cu,C)-1234, which is analogous to that of Tl-1234 and Hg-1234. Looking back to literatures, there is a controversial whether carbon is necessary in this compound since its discovery. Shimakawa et al. \cite{crystal structure} did a structure refinement for (Cu,C)-1234 using the neutron powder diffraction, and found that partial Cu sites in the charge reservoir layer could be substituted by the C atoms in the form of CO$_3$ groups. Recently, Duan et al. \cite{CuC film2} have confirmed the presence of CO$_3^{2-}$ carbonate clusters in the (Cu,C)-1234 thin films by the mid-infrared transmittance measurements and found that the content of CO$_3^{2-}$ carbonate clusters plays an important role on the superconducting properties of the films. The high-quality single crystal will provide another excellent platform for settling this issue.

The inset in Fig.~\ref{fig1}(b) presents the SEM image of the (Cu,C)-1234 single crystal, in which the layered structure and very flat surface morphology can be clearly observed. We also examine the chemical compositions of the single crystal by EDS analysis. Fig.~\ref{fig1}(b) displays a typical EDS spectrum for the single crystal as measured in the framed region in the inset, and the result is shown in Table~\ref{tbl}. The ratio of C: Ba: Ca: Cu is 0.31: 1.97: 3.0: 4.29 if we take Ca content as 3, which is very close to the stoichiometric composition of 0.4: 2.0: 3.0: 4.6 and confirms the formation of the (Cu,C)-1234 phase. If we move the excess content of Cu to the first atomic position, the atomic ratio in the crystal is 0.60: 1.97: 3.0: 4.0, namely (Cu$_{0.29}$C$_{0.31}$)Ba$_{1.97}$Ca$_3$Cu$_4$O$_{11+\delta}$. The composition by averaging the seven spots shown in the inset of Fig.~\ref{fig1}(b) is (Cu$_{0.31}$C$_{0.35}$)Ba$_{1.98}$Ca$_3$Cu$_4$O$_{11+\delta}$. Actually in the EDS analysis, the uncertainty about the carbon content content cannot be ignored. So we further measured the chemical composition of the single crystals by EPMA. The averaged results in one crystal show that the atomic ratio of C: Ba: Ca: Cu: O is 0.66: 1.65: 3.0: 4.23: 10.67, namely (Cu$_{0.23}$C$_{0.66}$)Ba$_{1.65}$Ca$_{3}$Cu$_4$O$_{10.67}$, and the carbon elements exist at all analyzed positions.

\begin{table}
  \caption{\label{tbl}Chemical compositions of the (Cu,C)-1234 single crystal examined by EDS measurements}
  \begin{center}
  \begin{tabular}{ccccccc}
    \hline
    \hline
     Element Symbol & Atomic Conc. ($\%$) & Weight Conc. ($\%$)\\
    \hline
    O & 69.97 & 34.84 \\
    Cu & 13.46 & 26.62 \\
    Ca & 9.41 & 11.73 \\
    Ba & 6.19 & 26.44 \\
    C & 0.97 & 0.36 \\
    \hline
    \hline
  \end{tabular}
  \end{center}
\end{table}

\begin{table}
  \caption{\label{tb2}Chemical compositions of the (Cu,C)-1234 single crystal determined by EPMA measurements}
  \begin{center}
  \begin{tabular}{ccccccc}
    \hline
    \hline
     Data & O (Mol$\%$) & Cu (Mol$\%$) & Ca (Mol$\%$) & Ba (Mol$\%$) & C (Mol$\%$)\\
    \hline
     Position 1 & 53.244 & 20.736 & 14.729 & 8.110 & 3.180\\
     Position 2 & 52.224 & 21.074 & 14.990 & 8.269 & 3.443\\
     Position 3 & 52.958 & 20.920 & 14.804 & 8.166 & 3.152\\
    \hline
    \hline
  \end{tabular}
  \end{center}
\end{table}

\begin{figure}
  \centering
  \includegraphics[width=4in]{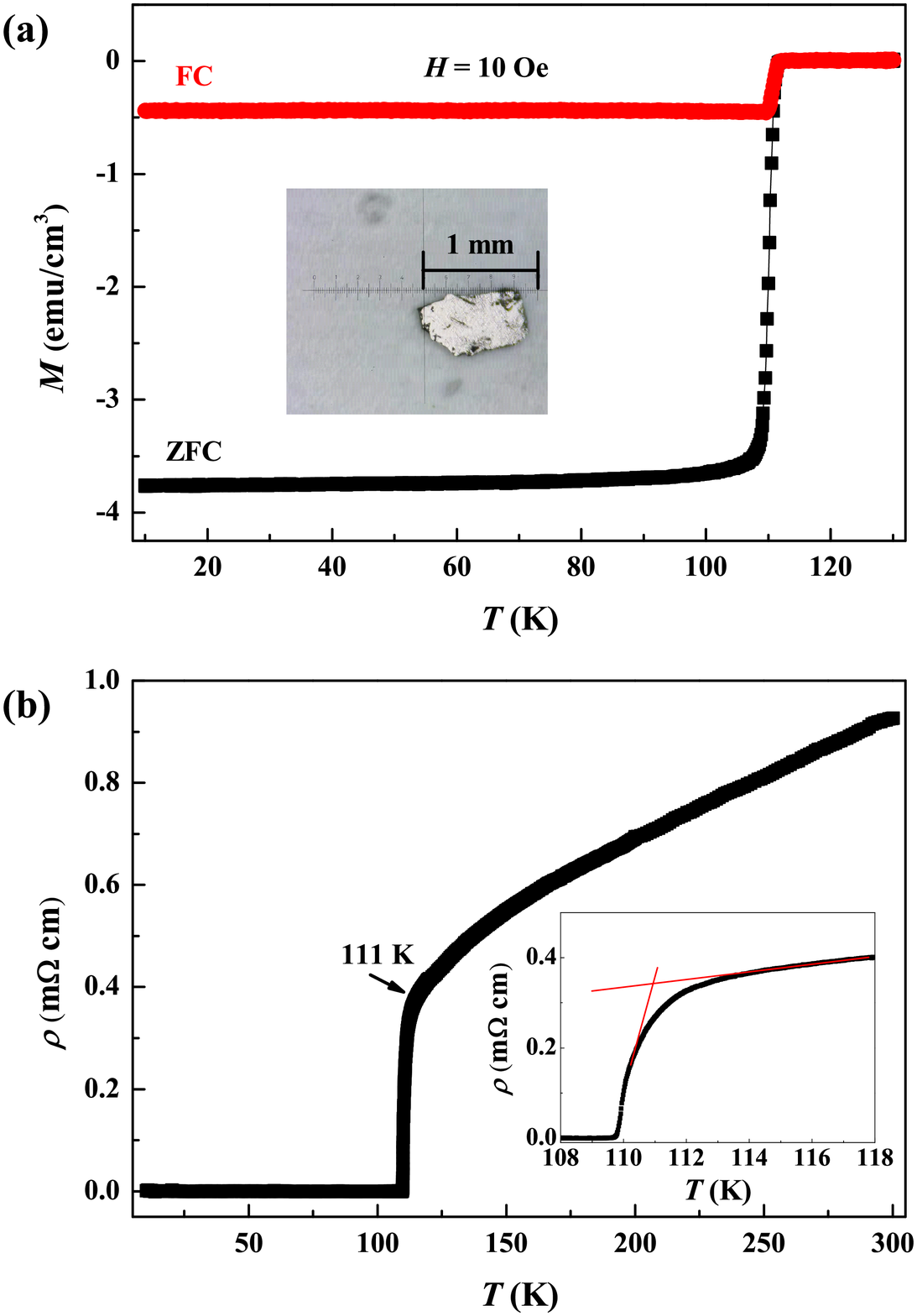}
\caption{\label{fig2}(a) Temperature dependence of magnetic susceptibility for the (Cu,C)-1234 single crystal measured in ZFC and FC modes at 10 Oe. The inset shows the photograph of the single crystal for the magnetic measurements with the crystal size of 0.9$\times$0.55$\times$0.1 mm$^3$. (b) Temperature dependence of resistivity for the (Cu,C)-1234 single crystal under zero magnetic field.}
\end{figure}

Figure~\ref{fig2}(a) shows the temperature dependence of magnetic susceptibility for the (Cu,C)-1234 single crystal measured in zero-field-cooled (ZFC) and field-cooled (FC) modes at 10 Oe. The inset shows the photograph of the single crystal for the magnetic measurements with the crystal size of 0.9$\times$0.55$\times$0.1 mm$^3$, which is significantly larger than that in Ref. \cite{Cu-1234 single crystal}. The \emph{M-T} curve displays a sharp superconducting transition with \emph{T}$_c$ of 112 K and the flat ZFC curve in the low temperature region indicates excellent diamagnetism. The shielding volume fraction at 10 K is estimated to be 472$\%$. The shielding volume beyond 100$\%$ is due to the demagnetization effect. Fig.~\ref{fig2}(b) presents the temperature dependence of resistivity for the (Cu,C)-1234 single crystal under zero magnetic field. The onset transition temperature is 111 K which is defined from the resistivity curve illustrated in the inset of Fig.~\ref{fig2}(b). The zero-resistance critical temperature is about 109.6 K, which is comparable to the value in the magnetic measurements. Both the magnetic and resistive measurements confirm the high quality of the single crystal. The \emph{T}$_c$ value in the present work is slightly lower than the value of 118 K in polycrystalline samples. In our experiments, we find that the \emph{T}$_c$ value in single crystals is related to the content of additive Ag$_2$O and the present sample is located in the slightly underdoped region. It is well known that the superconducting properties of cuprate superconductors are significantly influenced by the oxygen vacancies \cite{oxygen defect, oxygen}. In addition, the \emph{T}$_c$ value may also be influenced by the content of CO$_3^{2-}$ carbonate clusters, like the case in thin films \cite{CuC film2}.

\begin{figure}
  \centering
  \includegraphics[width=4in]{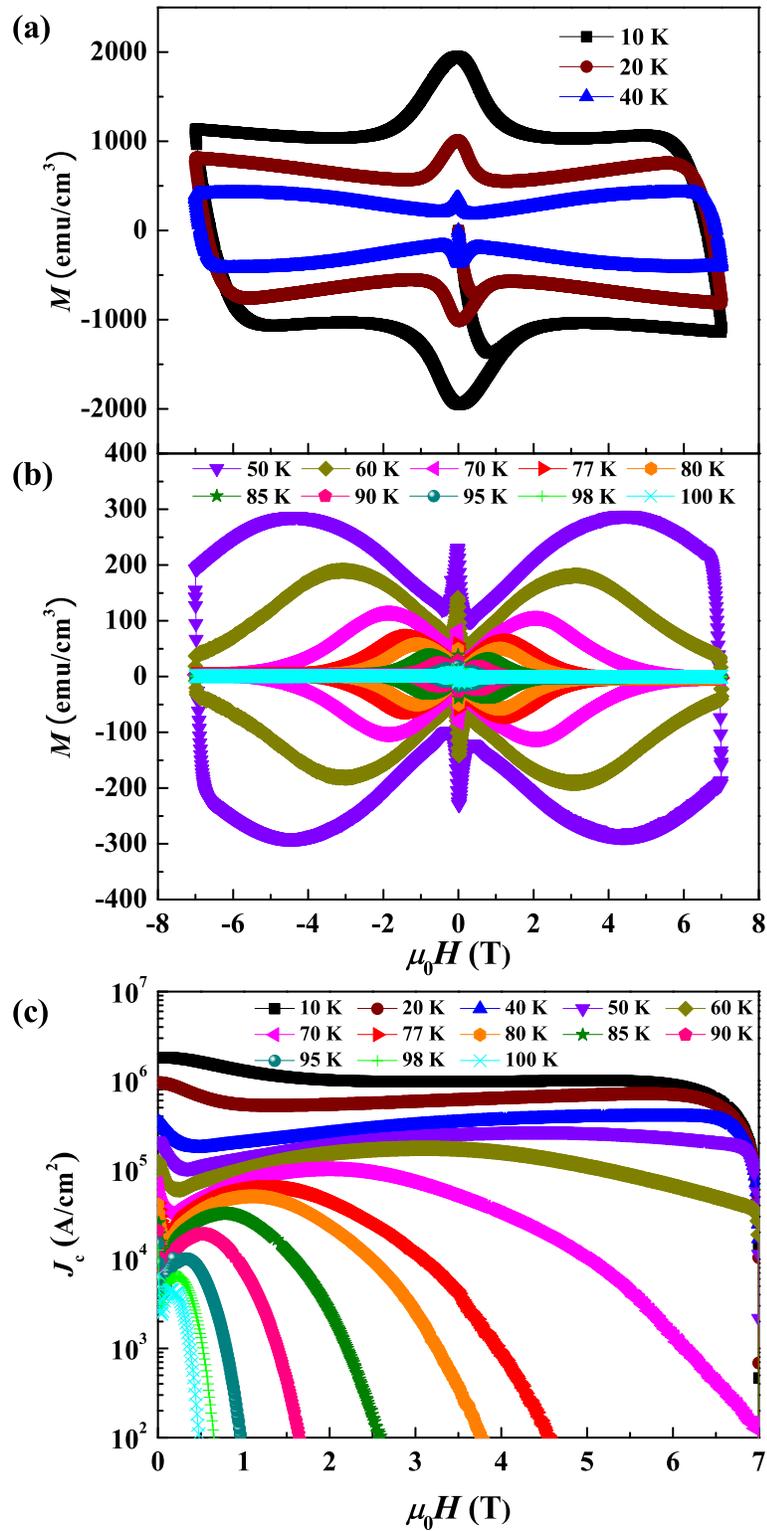}
\caption{\label{fig3}Magnetization hysteresis loops measured at temperatures from (a) 10 to 40 K and (b) 50 to 100 K with \emph{H}$\parallel$c. (c) Magnetic field dependence of critical current density \emph{J}$_c$ estimated by Bean critical state model.}
\end{figure}

In Figures~\ref{fig3}(a) and (b), we present the MHLs measured at different temperatures from 10 to 100 K with \emph{H}$\parallel$c. Interestingly, a pronounced second peak effect can be observed in Fig.~\ref{fig3}(b) at temperatures over 50 K. The peak shifts to higher fields with the decrease of temperature, and finally goes beyond 7 T, as shown in Fig.~\ref{fig3}(a). The second peak effect has also been observed in other superconductors and different mechanisms have been proposed to explain the origin according to the type of superconductors. In the low \emph{T}$_c$ superconductors such as Nb$_3$Sn \cite{Nb3Sn}, the second peak effect occurs in the vicinity of the upper critical field \emph{H}$_{c2}$, which is thought to be caused by the fast softening of the shear modulus C$_{66}$ of the flux line lattice \cite{low Tc spe}. While in high \emph{T}$_c$ superconductors, the second peak effect emerges at the field far from \emph{H}$_{c2}$ and can be classified into two types according to the temperature dependence of peak position. One is represented by Bi-2212 where the peak position locates at a low magnetic field, about 300-500 Oe, and is almost temperature independent \cite{Bi spe1}. It is interpreted as a consequence of the phase transition from the low-field ordered Bragg glass to the high-field disordered vortex glass \cite{Bi spe2, Bi spe3, Bi-2212 spe}. The other is represented by YBCO where the peak position is strongly temperature dependent \cite{YBCO spe1}. It has been demonstrated that the oxygen deficiencies can significantly affect the flux pinning and the second peak effect \cite{YBCO spe2}. And the inhomogeneity of oxygen distribution produces oxygen deficient regions which can act as extra pinning centers, and hence leads to the widening of MHLs in the high field region \cite{YBCO spe3}. The present result in (Cu,C)-1234 single crystal is similar to that in YBCO, which indicates that the second peak effect here may have the same origin as that in YBCO. One plausible explanation is that the (Cu,C)-1234 single crystal is more three dimensional and contains oxygen deficiencies which result in many local pinning centers.

The critical current density \emph{J}$_c$ can be calculated from MHLs by using the Bean critical state model \cite{Bean} with the formula \emph{J}$_c$ = 20$\triangle$\emph{M}/[a(1-a/3b)]. Here $\triangle$\emph{M} is the width of MHLs with the unit of emu/cm$^3$; a and b are the width and length of the sample with the unit of cm (b $\textgreater$ a), and the applied magnetic field is parallel to the c-axis of the single crystal. The results are displayed in Fig.~\ref{fig3}(c). The calculated \emph{J}$_c$ reaches 1.8$\times$10$^6$ A/cm$^2$ at 10 K and 0 T and 4.3$\times$10$^4$ A/cm$^2$ at 77 K and 0 T respectively, which are significantly higher than that in YBCO single crystals \cite{YBCO current}. Due to the presence of second peak effect, the \emph{J}$_c$ reaches 6.4$\times$10$^4$ A/cm$^2$ at 77 K and 1.5 T, which indicates good current carrying ability of the (Cu,C)-1234 single crystals.

\begin{figure}
  \centering
  \includegraphics[width=4in]{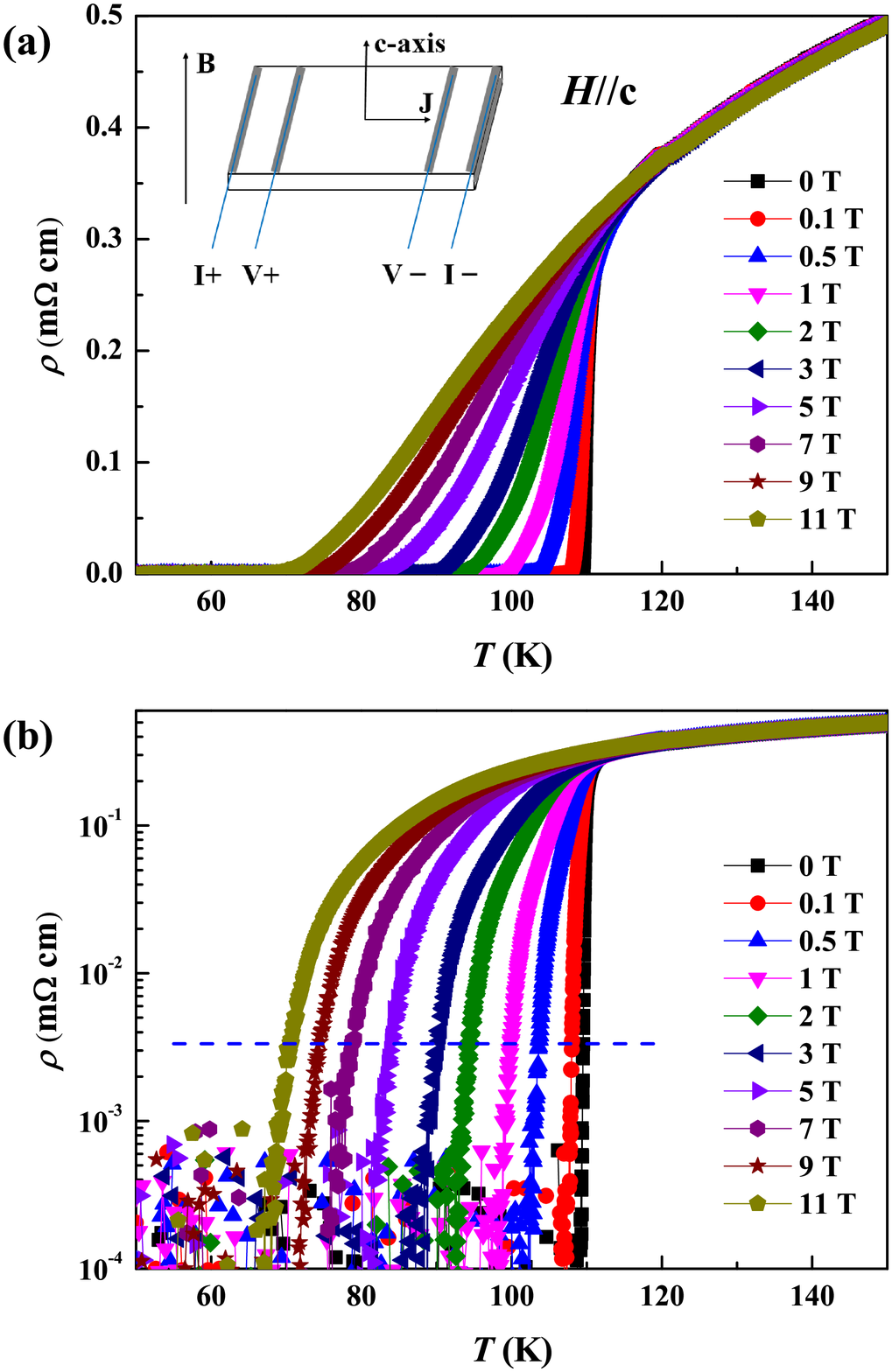}
\caption{\label{fig4}(a) Temperature dependence of resistivity at different fields from 0 to 11 T with \emph{H}$\parallel$c. (b) Temperature dependence of resistivity at different fields from 0 to 11 T with semilogarithmic scale. The blue horizontal dashed line represents the criterion of 1$\%$$\rho$$_n$, which is used to determine the irreversibility field.}
\end{figure}

In order to study the \emph{H}$_{irr}$ of the (Cu,C)-1234 single crystal, we performed the measurements of temperature dependence of resistivity at different magnetic fields with \emph{H}$\parallel$c, while the current is always applied in the ab plane (see the inset in Fig.~\ref{fig4}(a)). As shown in Fig.~\ref{fig4}(a), with the increase of the applied magnetic field, the superconducting transition is suppressed gradually. The zero-resistance critical temperature shifts to lower temperatures and the transition becomes broader with the increase of magnetic field, which is caused by the flux motion. By using a criterion of 1$\%$$\rho$$_n$ which has been illustrated by the blue dashed line in Fig.~\ref{fig4}(b), we can determine the \emph{H}$_{irr}$. The irreversibility fields are 11 T at 71 K, 8 T at 77 K and 3 T at 90 K. Furthermore, we have measured the temperature dependence of in-plane resistivity under different magnetic fields with \emph{H}$\parallel$c and \emph{H}$\parallel$ab for another piece of (Cu,C)-1234 single crystal. The superconducting anisotropy factor determined through the ratio of upper critical field \emph{H}$_{c2}^{ab}$(0)/\emph{H}$_{c2}^c$(0) is about 5.2.

\begin{figure}
  \centering
  \includegraphics[width=4in]{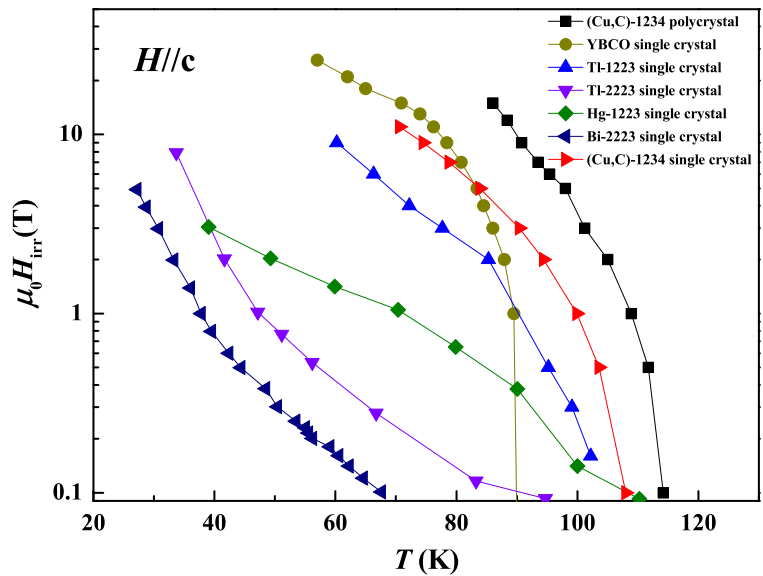}
\caption{\label{fig5}Irreversibility lines for (Cu,C)-1234 single crystal (this work), (Cu,C)-1234 polycrystal \cite{CuC final}, YBCO single crystal \cite{YBCO2}, Tl-1223 single crystal \cite{Tl-1223 IL}, Tl-2223 single crystal \cite{Tl-2223 IL2}, Hg-1223 single crystal \cite{Hg-1223 IL} and Bi-2223 single crystal \cite{Bi-based}.}
\end{figure}

Figure~\ref{fig5} shows the comparison of irreversibility lines for various cuprate systems with \emph{H}$\parallel$c, including (Cu,C)-1234 single crystal (this work), (Cu,C)-1234 polycrystal \cite{CuC final}, YBCO single crystal \cite{YBCO2}, Tl-1223 single crystal \cite{Tl-1223 IL}, Tl-2223 single crystal \cite{Tl-2223 IL2}, Hg-1223 single crystal \cite{Hg-1223 IL} and Bi-2223 single crystal \cite{Bi-based}. We can see that the \emph{H}$_{irr}$ at 77 K in the nontoxic Bi-2223 is very low due to the high anisotropy \cite{Bi-based}. The nontoxic YBCO single crystals have a higher \emph{H}$_{irr}$ of about 10 T at 77 K compared with Bi-2223. Furthermore, the nontoxic (Cu,C)-1234 polycrystal has the highest irreversibility field among all superconductors. The irreversibility fields of Tl-1223, Tl-2223 and Hg-1223 single crystals are lower than that of YBCO for \emph{T} $\textless$ 90 K, but higher for \emph{T} $\textgreater$ 90 K because of the higher \emph{T}$_c$ in the former three systems. However, the toxic elements Tl and Hg limit their practical applications. The irreversibility line of the (Cu,C)-1234 single crystal in the present work is lower than that of (Cu,C)-1234 polycrystal, which is reasonable considering the lower \emph{T}$_c$ of the single crystal. However, it is comparable to that in YBCO single crystal for \emph{T} $\textless$ 84 K and much higher for \emph{T} $\textgreater$ 84 K. And there is a large area between the irreversibility lines of YBCO single crystal and (Cu,C)-1234 single crystal for \emph{T} $\textgreater$ 84 K, which indicates that the (Cu,C)-1234 single crystal has greater potential for applications in this temperature region. Furthermore, the irreversibility line in (Cu,C)-1234 single crystal is higher than those in Tl-based and Hg-based systems. Thus, the (Cu,C)-1234 single crystal is expected to have great potential for applications in high fields with optimized \emph{T}$_c$ and improved properties.

\section{Conclusions}

In summary, millimeter-size (Cu,C)-1234 single crystals have been grown by high pressure and high temperature synthesis technique. The crystal structure, chemical composition, magnetization and resistivity measurements are carried out and all show good features of the single crystal. The \emph{T}$_c$ of about 111 K accompanied with a sharp superconducting transition is confirmed by both magnetic and resistive measurements. A pronounced second peak effect has been observed in both MHLs and \emph{J}$_c$(\emph{H}) curves. The single crystal exhibits a rather high critical current density and irreversibility line, which could be further improved with the optimized \emph{T}$_c$ of 118 K as previously discovered in polycrystalline samples. The present work will promote the fundamental research and application of the (Cu,C)-1234 material.

\section*{Acknowledgments}

This work was supported by the National Key R\&D Program of China (Grant Nos. 2016YFA0300401 and 2016YFA0401704), National Natural Science Foundation of China (Grant Nos. A0402/11534005 and E0209/52072170), and the Strategic Priority Research Program of Chinese Academy of Sciences (Grant No. XDB25000000).


\section*{References}

\begin{thebibliography}{00}
\bibitem{LaBaCuO}Bednorz J G and M\"{u}ller K A 1986 {\it Z. Phys. B: Condens. Matter} \textbf{64} 189-93.
\bibitem{Hirr}Nakane T, Fujinami K, Karppinen M and Yamauchi H 1999 {\it Supercond. Sci. Technol.} \textbf{12} 242-7.
\bibitem{Bi-based}Clayton N, Musolino N, Giannini E, Garnier V and Fl\"{u}kiger R 2004 {\it Supercond. Sci. Technol.} \textbf{17} S563-7.
\bibitem{Tl-based1}Liu R S, Zheng D N, Loram J W, Mirza K A, Campbell A M and Edwards P P 1992 {\it Appl. Phys. Lett.} \textbf{60} 1019-21.
\bibitem{Tl-based2}Zheng D N, Campbell A M, Liu R S and Edwards P P 1993 {\it Cryogenics} \textbf{33} 46-9.
\bibitem{Hg-based}Shirage P M, Iyo A, Shivagan D D, Crisan A, Tanaka Y, Kodama Y and Kito H 2008 {\it Physica C} \textbf{468} 1287-90.
\bibitem{YBCO1}Petrean A M, Paulius L M, Kwok W-K, Fendrich J A and Crabtree G W 2000  {\it Phys. Rev. Lett.}  \textbf{84} 5852-5.
\bibitem{YBCO2}Nishizaki T, Naito T and Kobayashi N 1998 {\it Phys. Rev. B} \textbf{58} 11169-72.
\bibitem{YBCO application}Larbalestier D, Gurevich A, Feldmann D M and Polyanskii A 2001 {\it Nature} \textbf{414} 368-77.
\bibitem{CuC-kawashima1994}Kawashima T, Matsui Y and Takayama-Muromachi E 1994 {\it Physica C} \textbf{224} 69-74.
\bibitem{CuC 118K}Crisan A, Badica P, Hirai M, Kito H, Iyo A and Tanaka Y 2002 {\it Supercond. Sci. Technol.} \textbf{15} 1240-3.
\bibitem{CuMg}Agarwal S K \emph{et al} 1998 {\it Phys. Rev. B} \textbf{58} 9504-9.
\bibitem{CuC anisotropy}Ihara H 2001 {\it Physica C} \textbf{364-365} 289-97.
\bibitem{CuC final}Zhang Y, Liu W, Zhu X, Zhao H, Hu Z, He C and Wen H-H 2018 {\it Sci. Adv.} \textbf{4} eaau0192.
\bibitem{CuC film1}Duan T, Hao J, Chu H and Wen H-H 2020 {\it Supercond. Sci. Technol.} \textbf{33} 025009.
\bibitem{defects}Zhu Y, Tafto J and Suenaga M 1991 {\it MRS Bull.} \textbf{16} 54-9.
\bibitem{YBCO current two times}Crabtree G W, Liu J Z, Umezawa A, Kwok W K, Sowers C H, Malik S K, Veal B W, Lam D J, Brodsky M B and Downey J W 1987 {\it Phys. Rev. B} \textbf{36} 4021-4.
\bibitem{IL high}Dubois S, Carmona F and Flandrois S 1996 {\it Physica C} \textbf{260} 19-24.
\bibitem{Cu-1234 single crystal}Tokiwa K \emph{et al} 1998 {\it Physica C} \textbf{298} 209-16.
\bibitem{crystal structure}Shimakawa Y, Jorgensen J D, Hinks D G, Shaked H, Hitterman R L, Izumi F, Kawashima T, Takayama-Muromachi E and Kamiyama T 1994 {\it Phys. Rev. B} \textbf{50} 16008-14.
\bibitem{CuC film2}Duan T, Hao J, Chu H, Li B, Dai Y and Wen H-H 2020 {\it Physica C} \textbf{573} 1353646.
\bibitem{oxygen defect}Daeumling M, Seuntjens J M and Larbalestier D C 1990 {\it Nature} \textbf{346} 332-5.
\bibitem{oxygen}Fratini M, Poccia N, Ricci A, Campi G, Burghammer M, Aeppli G and Bianconi A 2010 {\it Nature} \textbf{466} 841-4.
\bibitem{Nb3Sn}Lortz R, Musolino N, Wang Y, Junod A and Toyota N 2007 {\it Phys. Rev. B} \textbf{75} 094503.
\bibitem{low Tc spe}Pippard A B 1969 {\it Philos. Mag.} \textbf{19} 217-20.
\bibitem{Bi spe1}Yeshurun Y, Bontemps N, Burlachkov L and Kapitulnik A 1994 {\it Phys. Rev. B} \textbf{49} 1548-51.
\bibitem{Bi spe2}Giamarchi T and Le Doussal P 1994 {\it Phys. Rev. Lett.} \textbf{72} 1530-3.
\bibitem{Bi spe3}Giamarchi T and Le Doussal P 1995 {\it Phys. Rev. B} \textbf{52} 1242-70.
\bibitem{Bi-2212 spe}Li S and Wen H-H 2002 {\it Phys. Rev. B} \textbf{65} 214515.
\bibitem{YBCO spe1}Welp U, Kwok W K, Crabtree G W, Vandervoort K G and Liu J Z 1990 {\it Appl. Phys. Lett.} \textbf{57} 84-6.
\bibitem{YBCO spe2}Wen H-H and Zhao Z-X 1996 {\it Appl. Phys. Lett.} \textbf{68} 856-8.
\bibitem{YBCO spe3}Klein L, Yacoby E R, Yeshurun Y, Erb A, M\"{u}ller-Vogt G, Breit V and W\"{u}hl H 1994 {\it Phys. Rev. B} \textbf{49} 4403-6.
\bibitem{Bean}Bean C P 1964 {\it Rev. Mod. Phys.} \textbf{36} 31-9.
\bibitem{YBCO current}Choi K-Y, Jo I S, Han S-C, Han Y-H, Sung T-H, Jung M H, Park G S and Lee S-I 2011 {\it Curr. Appl. Phys.} \textbf{11} 1020-3.
\bibitem{Tl-1223 IL}Zheng D N, Johnson J D, Jones A R, Campbell A M, Liang W Y, Doi T, Okada M and Higashyama K 1995 {\it J. Appl. Phys.} \textbf{77} 5287-92.
\bibitem{Tl-2223 IL2}Brandst\"{a}tter G, Sauerzopf F M and Weber H W 1997  {\it Phys. Rev. B} \textbf{55} 11693-701.
\bibitem{Hg-1223 IL}Karpinski J, Angst M, Meijer G I, Kazakov S, Schwer H, Wisniewski A, Puzniak R 1999 {\it J. Low Temp. Phys.} \textbf{117} 735-9.

\end{thebibliography}
\end{document}